\documentclass[11pt,draftcls,onecolumn]{IEEEtran}

\usepackage{amsmath,url}
\usepackage{epsfig,amssymb,amsbsy,verbatim,array}
\usepackage{pstricks,psfrag,theorem,cite,enumerate}%

\let\intern=\iftrue

\def\argmax{\operatorname{arg~max}}
\def\figref#1{Fig.\,\ref{#1}}%
\def\E{\mathbb{E}}
\def\P{\mathbb{P}}
\def\R{\mathbb{R}}
\def\N{\mathbb{N}}

\def\ie{{\em i.e.}}
\def\eg{{\em e.g.}}
\def\V{\operatorname{Var}}

\def\erf{\operatorname{erf}}

\def\supp{\operatorname{supp}}
\def\d{\textnormal{d}}
\def\opt{\mathrm{opt}}

\def\one{\mathbf{1}}
\def\Ropt{{R_{\mathrm{opt}}}}
\def\W{\mathcal{W}}

\DeclareMathAlphabet{\mathsfsl}{OT1}{cmss}{m}{sl}
\def\x{\mathsfsl x}
\def\f{\mathsfsl f}

\def\y{\mathsfsl y}
\def\z{\mathsfsl z}
\def\r{\mathsfsl r}

\def\d{\text{d}}

\theoremstyle{plain}
\newtheorem{theorem}{Theorem}
\newtheorem{corollary}[theorem]{Corollary}

\newtheorem{proposition}[theorem]{Proposition}



\newlength{\figwidth}
\begin{document}

\title{A Geometric Interpretation of Fading in Wireless Networks: Theory
and Applications}
\author{Martin Haenggi,~\IEEEmembership{Senior Member,~IEEE}
\thanks{This paper is an extension of preliminary work that has appeared at ISIT 2006, Seattle, WA,
and ISIT 2007, Nice, France.
M.~Haenggi is with the Department~of
Electrical~Engineering, University of Notre Dame, Notre Dame, IN
46556, USA. E-mail: {\tt mhaenggi@nd.edu} }}
\maketitle
\begin{abstract}
In wireless networks with random node distribution, the underlying
point process model and the channel fading process are usually
considered separately. A unified framework is introduced that permits the
geometric characterization of fading by incorporating the fading
process into the point process model.
Concretely,  assuming nodes are distributed in a stationary Poisson
point process in $\R^d$, the properties of the point
processes that describe the path loss with fading are analyzed. The main applications are
connectivity and broadcasting. 
\end{abstract}

\section{Introduction and System Model}

\subsection{Motivation}
The path loss over a wireless link is well
modeled by the product of a distance component (often called
large-scale path loss) and a fading component (called small-scale
fading or shadowing). It is usually assumed that the distance part is
deterministic while the fading part is modeled as a random
process. This distinction, however, does not apply to many types of
wireless networks, where the distance itself is subject to uncertainty. In this
case it may be beneficial to consider the distance and fading
uncertainty jointly, \ie, to define a stochastic point process that incorporates
both. Equivalently, one may regard the distance uncertainty as
a large-scale fading component and the multipath fading uncertainty
as small-scale fading component.

We introduce a framework that offers such a geometrical
interpretation of fading and some new insight into its effect on the
network.  To obtain concrete analytical results, we will often use
the Nakagami-$m$ fading model, which is fairly general and offers the
advantage of including the special cases of Rayleigh fading and no fading
for $m=1$ and $m\rightarrow\infty$, respectively.

The two main applications of the theoretical foundations laid in
Section 2 are connectivity (Section 3) and
broadcasting (Section 4).

{\em Connectivity.} We characterize the geometric properties
of the set of nodes that are directly connected to the origin for arbitrary
fading models, generalizing the results in
\cite{net:Miorandi05,net:Haenggi06isit}. We also show that if
the path loss exponent equals the number of network dimension,
any fading model (with unit mean) is distribution-preserving in
a sense made precise later.

{\em Broadcasting.} We are interested in the single-hop {\em broadcast
transport capacity}, \ie, the cumulated distance-weighted rate summed
over the set of nodes that can successfully decode a message sent from
a transmitter at the origin. In particular, we prove that if the path
loss exponent is smaller than the number of network dimensions plus
one, this transport capacity can be made arbitrarily large
by letting the rate of transmission approach 0.

In Section 5, we discuss several other applications, including
the maximum transmission distance, probabilistic progress, the
effect of retransmissions, and localization.

\subsection{Notation and symbols}
For convenient reference, we provide a list of the
symbols and variables used in the paper. Most of them
are also explained in the text. Note that slanted sans-serif symbols
such as $\x$ and $\f$ denote random variables, in contrast to $x$ and $f$
that are standard real numbers or ``dummy" variables.
Since we model the distribution
of the network nodes as a stochastic point process, we use
the terms points and nodes interchangeably.

\[ 
\begin{array}{|c|l|}
\hline
\text{Symbol} & \text{Definition/explanation}\\\hline
 [ k ] & \text{the set }\{1,2,\ldots,k\} \\
 \one_A(x) & \text{indicator function}\\
u(x) & \triangleq \one_{\{x\geqslant 0\}}(x)\text{ (unit step function)}\\
d & \text{number of dimensions of the network}\\
o & \text{origin in }\R^d\\
B & \text{a Borel subset of }\R \text{ or }\R^d\\
c_d & \triangleq \pi^{d/2}/\Gamma(1+d/2)\\
        & \text{ (volume~of the $d$-dim.~unit ball)}\\
\alpha & \text{path loss exponent}\\
\delta & \triangleq d/\alpha\\
\Delta & \triangleq (d+1)/\alpha\\
s & \text{minimum path gain for connection}\\
 F,\f & \text{fading distribution (cdf), fading r.v.}\\
 F_X & \text{distribution of random variable }X\text{ (cdf)}\\
 \Phi=\{\x_i\} & \text{path loss process before fading (PLP)}\\
 \Xi=\{\xi_i\} & \text{path loss process with fading (PLPF)}\\
 \hat\Phi=\{\hat\x_i\} & \text{points in }\Phi\text{ connected to origin}\\
 \hat\Xi=\{\hat\xi_i\} & \text{points in }\Xi\text{ connected to origin}\\
 \Lambda,\lambda & \text{counting measure and density for }\Phi\\
 \hat N=\hat\Xi(\R^+) & \text{number of nodes connected to $o$}\\
\#A & \text{cardinality of }A\\
\hline
\end{array}
\]
~\\[-5mm]
\subsection{Poisson point process model}
A well accepted model for the node distribution in wireless networks\footnote{In
particular, if nodes move around randomly and independently, or if
sensor nodes are deployed from an airplane in large quantities.} is
the homogeneous \emph{Poisson point process} (PPP) of intensity $\lambda$.
Without loss of generality, we can assume $\lambda=1$
(scale-invariance).

{\em Node distribution.}
Let the set
$\{\y_i\}$, $i\in\N$ consist of the points of a stationary Poisson
point process in $\R^d$ of intensity $1$,
ordered according to their Euclidean distance $\|\y_i-o\|$ to the origin $o$.
Define a new one-dimensional
(generally inhomogeneous) PPP $\{\r_i\triangleq \|\y_i-o\|\}$ such that
$0<\r_1<\r_2<\ldots$ a.s.
Let $\alpha>0$ be the path loss exponent of the network and
$\Phi=\{\x_i\triangleq \r_i^\alpha \}$ be the {\em path loss process} 
(before fading) (PLP).
Let $\{\f,\f_1,\f_2,\ldots \}$ be an iid stochastic process with $\f$ drawn from a
distribution $F\triangleq F_{\f}$ with unit mean, \ie, $\E\f=1$, and $\supp\f \subset\R^+$.
Finally, let
$\Xi=\{\xi_i\triangleq \x_i/\f_i\}$ be the {\em path loss process with fading}
(PLPF). In order to treat the case of no fading in the same framework,
we will allow the degenerate case $F(x)=u(x-1)$, resulting in
$\Phi=\Xi$.
Note that the fading is static (unless mentioned otherwise),
and that $\{\xi_i\}$ is no longer ordered in general.
We will also interpret these point processes as
{\em random counting measures},
\eg, $\Phi(B)=\#\{\Phi\cap B\}$ for any Borel subset $B$ of $\R$.

{\em Connectivity.}  We are interested in connectivity to the origin.
A node $i$ is connected if its path loss is smaller than $1/s$, \ie,
if $\xi_i<1/s$. The processes of connected nodes are denoted as
$\hat\Phi=\{\x_i:\; \xi_i<1/s\}$ (PLP) and $\hat\Xi=\{\xi_i : \:\xi_i<1/s\}=\Xi\cap[0,1/s)$ (PLPF).

{\em Counting measures.}
Let $\Lambda$ be the counting measure associated with
$\Phi$, \ie, $\Lambda(B)=\E\Phi(B)$ for Borel $B$.
For $\Lambda([0,a))=\E\Phi([0,a))$, we will also use
the shortcut $\Lambda(a)$. 
Similarly, let $\hat\Lambda$ be the counting measure for
$\hat\Phi$.
All the point processes considered admit a {\em density}. Let
$\lambda(x)=\d\Lambda(x)/\d x$ and
and $\hat\lambda(x)=\d\hat\Lambda(x)/\d x$ be the densities
of $\Phi$ and $\hat\Phi$, respectively.

{\em Fading model.}
To obtain concrete results, we frequently use the Nakagami-$m$ (power) fading model.
The distribution and density are
\begin{align}
 F(x)&=1-\frac{\Gamma_\text{ic}(m,mx)}{\Gamma(m)}\\
 f(x)&=\frac{m^m x^{m-1}\exp(-m x)}{\Gamma(m)}\,,
\end{align}
where $\Gamma_\mathrm{ic}$
denotes the upper incomplete gamma function.
This distribution is a single-parameter version of the gamma
distribution where both parameters are the same such that
the mean is $1$ always.

\subsection{The standard network}
For ease of exposition, we often consider
a {\em standard network}\footnote{The term ``standard'' here refers
to the fact that in this case the analytical expressions are particularly
simple. We do not claim that these parameters are the ones
most frequently observed in reality.}
 that has the following parameters:
$\delta\triangleq d/\alpha=1$ (path loss exponent equals the number of dimensions) and
Rayleigh fading, \ie, $F(x)=(1-e^{-x})u(x)$.

\begin{figure}
\centerline{\epsfig{file=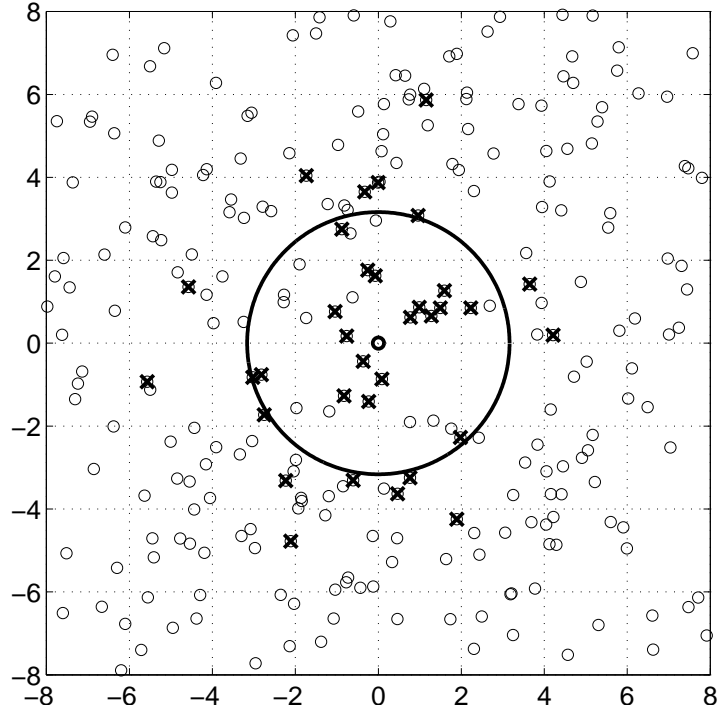,width=\figwidth}}
\caption{A Poisson point process of intensity 1 in a $16\times 16$ square. The reachable nodes
by the center node are indicated by a bold $\times$ for a path gain
threshold of $s=0.1$, a path loss exponent of $\alpha=2$, and Rayleigh
fading (standard network). The circle indicates the range of successful
transmission in the non-fading case. Its radius is
$1/\sqrt{s}\approx 3.16$, and there are about $\pi/s\approx 31$ nodes inside.}
\label{fig:disks2}
\end{figure}

\figref{fig:disks2} shows a PPP of intensity 1 in a $16\times 16$ square,
with the nodes marked that can be reached from the center, assuming
a path gain threshold of $s=0.1$.
The disk shows the maximum transmission distance in the non-fading case.

\section{Properties of the Point Processes}
\begin{proposition}
The processes $\Phi$, 
$\Xi$, and $\hat\Xi$ are Poisson.
\end{proposition}
\begin{IEEEproof}
$\{y_i\}$ is Poisson by definition, so $\{r_i\}$ and $\Phi=\{x_i\}$
are Poisson by the mapping theorem \cite{net:Kingman93}. 
$\Xi$ is Poisson since $\f_i$ is iid, and $\hat\Xi(\R)=\Xi([0,1/s))$.
\end{IEEEproof}
The Poisson property of $\hat\Phi$ will be established in Prop.~\ref{prop:con}.

Cor.~\ref{cor:basic} states some basic facts about these point processes
that result from their Poisson property.

\begin{corollary}[Basic properties.]~\\[-6mm]
\label{cor:basic}
\begin{enumerate}[(a)]
\item $\Lambda(x)=\E \Phi([0,x))=c_d x^\delta$ and
  $\lambda(x)=c_d\delta x^{\delta-1}$. In particular,  for $\delta=1$,
  $\Phi$ is stationary (on $\R^+$).
\item $\r_i$ is governed by the generalized gamma pdf
\begin{equation}
f_{\r_i}(r)= e^{- c_d r^d}  \frac{d\:( c_d r^d)^i}{r\Gamma(i)}\,,
\label{weibull}
\end{equation}
and $\x_i$ is distributed according to the cdf
\begin{equation}
 F_{\x_i}(x)=1-\frac{\Gamma_\mathrm{ic}(i,c_d x^\delta)}{\Gamma(i)}\,,.
 \label{Fxi}
\end{equation}
The expected path loss without fading is
\begin{equation}
\E\x_i=c_d^{-1/\delta} \frac{\Gamma(i+1/\delta)}{\Gamma(i)}\,.
\end{equation}
In particular, for the standard network, the $x_i$ are Erlang with $\E x_i=i/c_d$.
\item The distribution function of $\xi_i$ is 
\begin{equation}
F_{\xi_i}(x)=1-\int_0^\infty F(r/x) \left(\frac{c_d^i \delta r^{\delta i-1} \exp(-c_d r^\delta)}
{\Gamma(i)}\right){\mathrm d} r\,.
\label{Fxii}
\end{equation}
For $\delta=1$ and Nakagami-$m$ fading, the pdf of $\xi_i$ is
\begin{equation}
  f_{\xi_i}(x)=\frac{m^{m+1}\binom{m+i-1}{m}c_d^i x^{i-1}}{(m+c_d x)^{m+i}}\,.
  \label{fxi_naka}
\end{equation}
In particular,
\begin{equation}
  F_{\xi_1}(x)=1-\left(\frac{m}{c_d x+m}\right)^m\,
  \label{fxi_naka1}
\end{equation}
and
\begin{align}
 \E \xi_i&=\frac{mi}{c_d(m-1)}\qquad\text{for }m>1\\
 \V \xi_i&=\frac{m^2 i(m+i-1)}{c_d^2(m-1)^2(m-2)}\qquad\text{for }m>2\,.
\end{align}
For the standard networks,
\begin{equation}
 F_{\xi_i}(x)=\left(\frac{c_d x}{c_d x+1}\right)^i\,.
\end{equation}
\end{enumerate}
\end{corollary}
\begin{IEEEproof}
\begin{enumerate}[(a)]
\item Since the original $d$-dimensional process $\{\y_i\}$ is stationary, the
  expected number of points in a ball of radius $x$ around
    the origin is $c_d x^d$. The one-dimensional process $\{\r_i\}$
    has the same number of points in $[0,x)$, and $\x_i=\r_i^\alpha$,
    so $\E\Phi([0,x))=c_d x^{\delta}$. For $\delta=1$, $\lambda(x)=c_d$
    is constant.
\item Follows directly from the fact that $\{\y_i\}$ is stationary Poisson.
    (\eqref{weibull} has been established in \cite{net:Haenggi03it}.)
\item The cdf $\P[\xi_i<x]$ is $1-\E_{\x_i}(F(\x_i/x))$ with $\x_i$ distributed according to  \eqref{Fxi}.
 \eqref{fxi_naka} is obtained by straightforward (but tedious) calculation.
 \end{enumerate}
\end{IEEEproof}
{\em Remarks:} 
\begin{itemize}
\item[-] For general (rational) values of $m$,  $d$, and $\alpha$, $F_{\xi_i}$ can be expressed
    using hypergeometric functions.
\item[-]
  \eqref{fxi_naka1} approaches $1-\exp(-c_d x)$ as $m\rightarrow\infty$, which
  is the distribution of $x_1$. Similarly, $\lim_{m\rightarrow\infty} \E\xi_i=i/c_d=\E\x_i$ and
  $\lim_{m\rightarrow\infty} \V\xi_i=i/c_d^2=\V\x_i$.
\item[-] Alternatively we could consider the {\em path gain process} $\xi_i^{-1}$.  Since
    $F_{\xi_i^{-1}}(x)=1-F_{\xi_i}(1/x)$, the distribution functions look similar.
\item[-] In the standard network, the expected path loss $\E \xi_i$ does not exist for any $i$,
   and for $i=1$,  the expected path gain is infinite, too, since both $x_1$ and $\f$ are
   exponentially distributed. For $i>1$, $\E(\xi_i^{-1})=c_d/(i-1)$, and for $i>2$,
   $\V(\xi_i^{-1})=2c_d^2/((i-1)(i-2))$.
\item[-] For the standard network, the differential entropy $h(\xi_i)\triangleq\E[-\ln f_{\xi_i}(\xi_i)]$ is
  $2-\log c_d$ for $i=1$ and grows logarithmically with $i$. For Nakagami-$m$ fading
  $h(\xi_1)=1+1/m-\log c_d$. For the path gain process in the standard network, the 
  entropy has the simple expression
  \begin{equation}
     h(\xi_i^{-1})=\frac{i+1}{i}+\log\left(\frac{\pi}{i}\right)\,,
  \end{equation}
  which is monotonically {\em decreasing}, reflecting the fact that the variance
  $\V \xi_i^{-1}$ is decreasing with $i^{-2}$.
  \item[-] The $\xi_i$ are not independent since the $\x_i$ are ordered.
   For example, in the case of the standard
   network, the difference $\x_{i+1}-\x_i$ is exponentially distributed with mean $1/c_d$,
   thus the joint pdf is
   \begin{equation}
  f_{\x_1\ldots\x_n}(x_1,\ldots,x_n)=c_d^n e^{-c_d x_n}
   \mathbf{1}_{0< x_1<\ldots <x_n}\,,
\end{equation}
where $\mathbf{1}_{0< x_1<\ldots <x_n}$ denotes the (positive)
order cone (or hyperoctant) in $n$ dimensions.
\end{itemize}

\begin{proposition}
For $\delta=1$ and any fading distribution $F$ with mean $1$,
\[ \Xi(B)\stackrel{d}{=}\Phi(B) \qquad\forall B\subset \R^+ \,,\]
i.e.,
fading is distribution-preserving.
\end{proposition}
\begin{IEEEproof}
 Since $\Xi$ is Poisson, independence of
 $\Xi(B_1)$ and $\Xi(B_2)$ for $B_1\cap B_2=\emptyset$
 is guaranteed. So it remains to be shown that the
 intensities (or, equivalently, the counting
 measures on Borel sets) are the same. This is the case if
 for all $a>0$, 
 \[ \E\left(\#\{\x_i:  \x_i > a ,\: \xi_i < a \}\right)=\E\left(\#\{\x_i : \x_i<a ,\: \xi_i >a\}\right)\,, \]
   \ie, the expected numbers of nodes crossing $a$ from the left (leaving
   the interval $[0,a)$) and
   the right (entering the same interval) are equal. This condition can be expressed as
 \[ \int_0^a \lambda(x)F (x/a)\d x = \int_a^\infty \lambda(x) (1-F(x/a)) \d x\quad\forall a>0\,.\]
 If $\delta=1$, $\lambda(x)=c_d$, and the condition reduces to
 \[ \int_0^1 F(x)\d x=\int_1^\infty (1-F(x))\d x \,,\]
 which holds since
 \[ \underbrace{\int_0^1 (1-F(x))\d x}_{1-\int_0^1 F(x)\d x}+\int_1^\infty (1-F(x))\d x
  =\E \f=1 \,. \]
\end{IEEEproof}
 
An immediate consequence is that a receiver cannot decide on the amount of
fading present in the network if $\delta=1$ and geographical distances are not known.
 
 \begin{corollary}
 For Nakagami-$m$ fading, $\delta=1$,
 and any $a>0$, the expected number of nodes with $x_i<a$ and $\xi_i>a$, \ie, nodes
 that leave the interval $[0,a)$ due to fading, is
\begin{equation}
    \E\left(\#\{\x_i : \x_i<a ,\: \xi_i >a\}\right)=c_d a \frac{m^{m-1}}{\Gamma(m)}e^{-m}\,. 
\end{equation}
The same number of nodes is expected to enter this interval.
For Rayleigh fading ($m=1$), the fraction of nodes leaving any
interval $[0,a)$ is $1/e$.
 \end{corollary}
 \begin{IEEEproof}
   $\E\left(\#\{\x_i : \x_i<a\}\right)=\Lambda(a)=c_d a$, and for Nakagami-$m$, the
   fraction of nodes leaving the interval is
   \[ \int_0^1 F(x)\d x=
   \frac{m^{m-1}}{\Gamma(m)}e^{-m}\,. \]
\end{IEEEproof}

\begin{figure}
\centerline{\epsfig{file=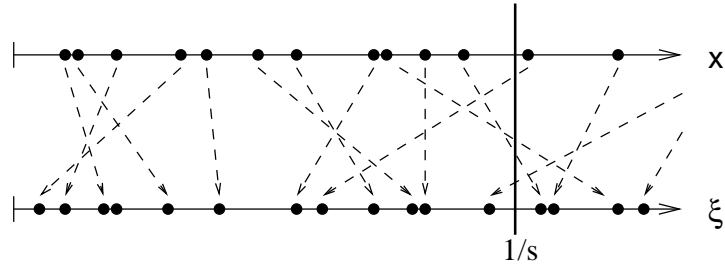,width=\figwidth}}
\caption{The points of a Poisson point process $\x_i$ are mapped
and reordered according to $\xi_i:=\x_i/\f_i$, where $\f_i$ is iid exponential
with unit mean.
In the lower axis, the nodes to the left of the threshold $1/s$ are
connected to the origin (path loss smaller than $1/s$).}
\label{fig:pathloss}
\end{figure}

Clearly, fading can be interpreted as a stochastic mapping from
$\x_i$ to $\xi_i$.
So, $\{\x_i\}$ are the points in the geographical domain (they indicate
distance), whereas $\{\xi_i\}$ are the points in the path loss domain,
since $\xi_i$ is the actual path loss including fading. This mapping
results in a partial reordering of the nodes, as
visualized in \figref{fig:pathloss}. In the path loss domain, the
connected nodes are simply given by $\{\hat\xi_i\}=\{\xi_i\}\cap [0,1/s]$.

\begin{figure}
\centerline{\epsfig{file=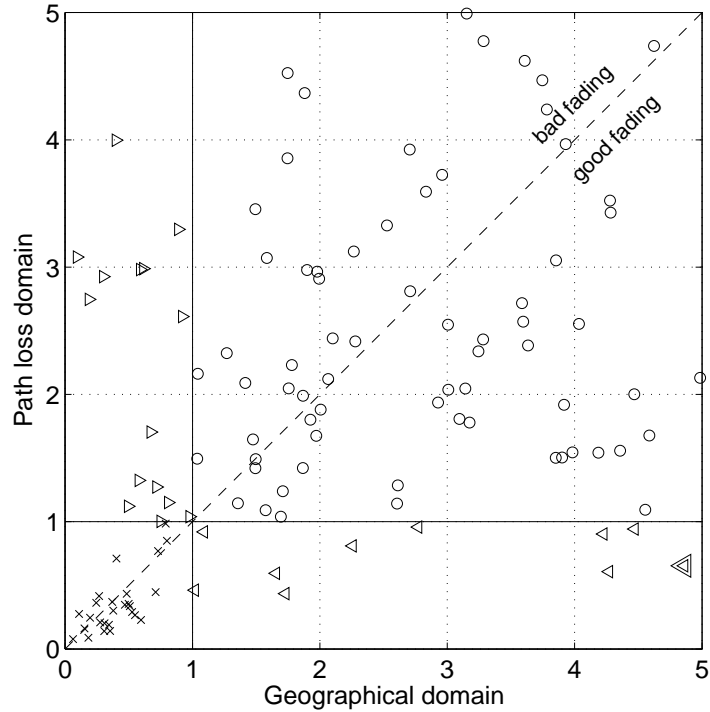,width=\figwidth}}
\caption{Illustration of the Rayleigh mapping. 200 points $x_i$ are
chosen uniformly randomly in $[0,5]$. Plotted are the points
$(x_i,x_i/f_i)$, where the $f_i$ are drawn iid exponential with mean 1. Consider the
interval $[0,1]$ (\ie, assume a threshold $s=1$). Points marked by
$\times$ are points that remain inside $[0,1]$, those marked by o
remain outside, the ones marked with left- and right-pointing
triangles are the ones that moved in and out, respectively. The node
marked with a double triangle is the furthest reachable node. On
average the same number of nodes move in and out. Note that not all
points are shown, since a fraction $e^{-1}$ is mapped outside of
$[0,5]$.}
\label{fig:mapping}
\end{figure}

\figref{fig:mapping} illustrates the situation for 200 nodes randomly
chosen from $[0,5]$ with a threshold $s=1$. Before fading, we expect
40 nodes inside. From these, a fraction $e^{-1}$ is moving out (right
triangles), the rest stays in (marked by $\times$). From the ones
outside, a fraction $(1-e^{-4})(ae)\approx 9\%$ moves in (left
triangles), the rest stays out (circles). 

For the standard network, the probability of point reordering due to fading can be calculated
explicitly.
Let $P_{i,j}\triangleq\P[\xi_i>\xi_{i+j}]$. By this definition,
\begin{equation}
  P_{i,j}=\P[\x_i/\f_i > \x_{i+j}/\f_{i+j}]=\P\left[\frac{\x_i}{\x_i+\y_j} > \frac{\f_i}{\f_{i+j}}\right]\,.
\end{equation}
$\x_i$ is Erlang with parameters $i$ and $c_d$,
$\y_j$ is the distance from $\x_i$ to $\x_{i+j}$ and thus Erlang with
parameters $j$ and $c_d$, and the
cdf of $\z:=\f_n/\f_{n+m}$ is $F_\z(x)=x/(x+1)$. Hence
\begin{align*}
 P_{i,j}\!=&\E_{\x,\y}\left(\frac{\x_i}{2\x_i+\y_j}\right)\\
  =&\int_0^\infty \!\!\int_0^\infty \frac{x}{2x+y}
  \frac{c_d^{i+j}x^{i-1}y^{j-1}}{\Gamma(i)\Gamma(j)}e^{-c_d(x+y)} \text{d}x\text{d}y\,.
\end{align*}
$P_{i,j}$ does not depend on $c_d$.
Closed-form expressions include
$P_{1,1}=1-\ln 2\approx 0.307$, and $P_{1,2}=3-4\ln 2\approx 0.227$.
Generally $P_{k,k}$ can be determined analytically. For $k=1,2,3,4$,
we obtain $1-\ln 2,\;12\ln 2-8,\;167/2-120\ln 2,\;1120\ln 2-776$. 
Further, $\lim_{k\rightarrow\infty}
P_{k,k}=1/3$, which is the probability that an exponential random variable
is larger than another one that has twice the mean.

In the limit, as $i\rightarrow\infty$, $P_{i,j}=1/(j+1)$,
which is the probability that a node has the largest fading coefficient
among $j+1$ nodes that are at the same distance. Indeed, as
$i\rightarrow\infty$, $\x_{i+j}<\x_i(1+\epsilon)$ 
a.s.~for any $\epsilon>0$ and finite $j$.\\

While the $\xi_i$ are dependent,
it is often useful to consider a set of {\em independent}
random variables, obtained by conditioning the process on having
a certain number of nodes $n$ in an interval $[0,a)$ (or,
equivalently, conditioning on $\x_{n+1}=a$) and randomly permuting
the $n$ nodes. In doing so, the $n$ points $\{\x_i\}$ and $\{\xi_i\}$, $i=1,2,\ldots,n$
are iid distributed as follows.

\begin{corollary}
Conditioned on $\x_{n+1}=a$:
\begin{enumerate}[(a)]
\item  The nodes $\{\x_i\}_{i=1}^n$ are iid
distributed with
\begin{equation}
  f_{\x_i}^a(x)=\frac{\lambda(x)}{\Lambda(x)}=\delta \left(\frac x a\right)^{\delta} \frac 1 x\,,
  \quad 0\leqslant x<a\,
  \label{fxia}
\end{equation}
and cdf $F_{\x_i}^a(x)=(x/a)^\delta$.
\item The path loss with fading $\{\xi_i\}_{i=1}^n$ is distributed as
\begin{equation}
F_{\xi_i}^a(x)=1-\int_0^a F(y/x) \delta \left(\frac y a\right)^{\delta} \frac 1 y\d y\,.
\label{Fxiia}
\end{equation}
\item For the standard network,
\begin{equation}
  F_{\xi_i}^a(x)=\frac x a \left(1-e^{-a/x}\right)\,
\end{equation}
\item For Rayleigh fading and $\delta=1/2$, 
\begin{equation}
F_{\xi_i}^a(x)=\frac{\sqrt\pi}2 \sqrt{\frac x a} \erf\left(\sqrt\frac a x\right)\,.
\end{equation}
\end{enumerate}
\end{corollary}
\begin{IEEEproof}
  As in \eqref{Fxii}, the cdf is given by $1-\E(F(\y/x))$ with $\y$ distributed
  as \eqref{fxia}.
\end{IEEEproof}

\section{Connectivity}
Here we investigate the processes $\hat\Phi$ and $\hat\Xi=\Xi\cap[0,1/s)$
of connected nodes. 

\subsection{Single-transmission connectivity and fading gain}

\begin{proposition}[Connectivity]
Let a transmitter situated at the origin transmit a single message, and
assume that nodes with path loss smaller than $1/s$ can decode, \ie,
are connected.
We have:
\label{prop:con}
\begin{enumerate}[(a)]
  \item $\hat\Phi$ is Poisson with $\hat\lambda(x)=\lambda(x)(1-F(sx))$.
  \item With Nakagami-$m$ fading, the number
  $\hat N=\hat\Phi(\R^+)$ of connected nodes is
  Poisson with mean
  \begin{equation}
  \E \hat N_m=\frac{c_d}{(ms)^\delta} \frac{\Gamma(\delta+m)}{\Gamma(m)}
  \label{hatN}
\end{equation}
and the {\em connectivity fading gain}, defined as the
ratio of the expected numbers of connected nodes with and
without fading, is
\begin{equation}
\frac{\E \hat N_m}{\E \hat N_\infty}= 
\frac 1{m^\delta} \frac{\Gamma(\delta+m)}{\Gamma(m)}=\E(\f^\delta)\,.
\label{con-fading-gain}
\end{equation}
\end{enumerate}
\end{proposition}
\begin{proof}
\begin{enumerate}[(a)]
\item The effect of fading on the connectivity is independent (non-homogeneous) thinning
  by $1-F(sx)=\P[x/\f<1/s]$.
\item Using (a), the expected number of connected nodes is 
\[ \int_0^\infty \hat\lambda(x)\d x=\int_0^\infty c_d\delta x^{\delta-1}
\frac{\Gamma_\mathrm{ic}(m,msx)}{\Gamma(m)} \d x\,\]
which equals $\E\hat N_m$ in the assertion.
Without fading, $\E\hat N_\infty=
\lim_{m\rightarrow\infty}=\Lambda(1/s)=c_d s^{-\delta}$, which
results in the ratio \eqref{con-fading-gain}.
\end{enumerate}
\vspace*{-5mm}
\end{proof}

\begin{figure}
\centerline{\epsfig{file=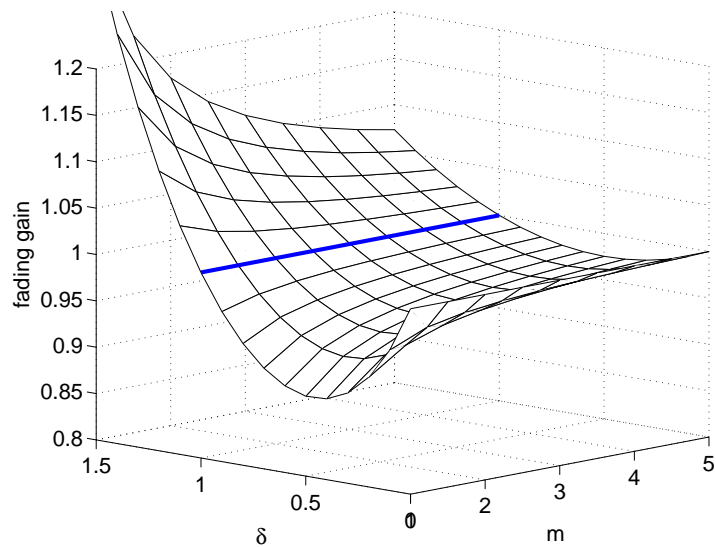,width=\figwidth}}
\caption{Connectivity fading gain for Nakagami-$m$ fading
 as a function of $\delta\in [0,3/2]$ and $m\in [1,5]$. 
For $\delta=1$, the gain is $1$ independent of $m$ (thick line).}
\label{fig:gains}
\end{figure}

{\em Remarks:}
\begin{enumerate}
\item \eqref{hatN} is a generalization of a result in \cite{net:Miorandi05} where the connectivity of
 a node in a two-dimensional network with Rayleigh fading was studied.
\item $\E \hat N$ can also be expressed as 
\begin{equation}
\E \hat N=\sum_{i=1}^\infty \P[\xi_i<1/s]\,.
\end{equation}
The relationship with part (b) can be viewed as a simple instance of Campbell's
theorem \cite{net:Stoyan95}. Since $\hat N$ is Poisson, the probability of isolation is
 $\P(\hat N=0)=\exp(-\E \hat N)$.
\item $\E \hat N_1=c_d s^{-\delta} \Gamma(\delta+1)$, and $\E \hat N_{\infty}=c_d s^{-\delta}$.
 For $\delta=1$, $\hat N$ does not depend on the type (or presence) of fading.
\item The {\em connectivity fading gain} equals the $\delta$-th moment
of the fading distribution, which, by definition, approaches one as the
fading vanishes, \ie, as $m\rightarrow\infty$. For a fixed $\delta$,
it is decreasing in $m$ if $\delta>1$, increasing if $\delta<1$, and
equal to $1$ for all $m$ if $\delta=1$. It also equals $1$ if $\delta=0$.
For a fixed $m$, it is not monotonic with $\delta$,
but exhibits a minimum at some $\delta_{\min}\in (0,1)$. The fading gain as a function
of $\delta$ and $m$ is plotted in \figref{fig:gains}.
For Rayleigh fading and
$\delta=1/2$, the fading gain is $\pi/2$, and the minimum is assumed at
$\delta_{\min}\approx 0.462$, corresponding to $\alpha\approx 4.33$ for $d=2$.
So, depending on the type of fading and the
ratio of the number of network dimensions to the path loss exponent
$\alpha$, fading can increase or decrease the number of connected nodes.
\item For the standard network, $\E\hat N=c_d/s$ and the probability of
isolation is $e^{-c_d/s}$.
\item 
The expected number of connected nodes $\hat N^a$ with
$\x_i<a$ is
\begin{equation}
\E \hat N^a=c_d a^\delta F_{\xi_i}^a(1/s)\,.
\end{equation}
where $F_{\xi_i}^a$ is given in \eqref{Fxiia}.
\end{enumerate}

\begin{corollary}
Under Nagakami-$m$ fading, 
a uniformly randomly chosen connected node $\hat\x\in\hat\Phi$ has mean
\begin{equation}
\E\hat\x=\frac{\delta(\delta+m)}{m s (\delta+1)}\,,
\end{equation}
which is $1+\delta/m$ times the value without fading.
\end{corollary}
\begin{proof}
A random connected node $\hat\x$ is distributed according to
\begin{equation}
  f_{\hat\x}(x)=\frac{\hat\lambda(x)}{\E\hat N}\,.
  \label{randomconnected_pdf}
\end{equation}
Without fading, the distribution is $s^\delta \delta x^{\delta-1}$,
$0\leqslant x\leqslant 1/s$, resulting in an expectation of
$\delta/(s(\delta+1))$.
\end{proof}

For Rayleigh fading, for example, the density
$f_{\hat x}$ is a gamma density with mean $\delta/s$, so
the average connected node is $1+\delta$ times further away than
without fading.

\subsection{Connectivity with retransmissions}
Assuming a block fading network and $n$ transmissions
of the same packet, what is the process of nodes that
receive the packet at least once?
\begin{corollary}
\label{cor:retrans}
In a network with iid block fading, the density of the 
process of nodes
$\hat\lambda^n$ that receive at least one of $n$ transmissions
is
\begin{equation}
\hat\lambda^n(x)= (1-F(sx)^n) c_d \delta x^{\delta-1}\,.
\end{equation}
\end{corollary}
\begin{IEEEproof}
  This is a straightforward generalization of Prop.~\ref{prop:con}(a).
\end{IEEEproof}
So, in a standard network, the number of connected nodes with
$n$ transmissions 
\begin{equation}
\E\hat N^n=\int_0^\infty \hat\lambda^n(x)\d x=\frac{c_d}s(\Psi(n+1)+\gamma)\,,
\label{conn_n}
\end{equation}
where $\Psi$ is the digamma function (the logarithmic derivative of the gamma
function),  which grows with $\log n$.
Alternatively if the threshold $s_k$ for the $k$-th transmission is
chosen as $s_k\triangleq s_1/k$, $k\in [n]$,
the expected number of nodes reached increases linearly with the
number of transmissions.

\section{Broadcasting}
\subsection{Broadcasting reliability}

\begin{proposition}
\label{prop:broad}
For $\delta=1$ and Nakagami-$m$ fading, $m\in\mathbb{N}$, the probability that a randomly chosen
node $\x\in [0,a)$ can be reached is
\begin{equation}
 p_m(\tilde s)=\frac 1{\tilde s}\left(1-\exp(-m\tilde s)\sum_{k=0}^{m-1} \frac{m^k(1-k/m)}{k!} \tilde{s}^k\right)\,,
\end{equation}
where $\tilde s\triangleq as$.
$p_m$ is increasing in $m$ for all $\tilde s>0$ and
converges uniformly to
\begin{equation}
  \lim_{m\rightarrow\infty} p_m(\tilde s)=\min\{1,\tilde{s}^{-1}\}\,.
\end{equation}
\end{proposition}
\begin{IEEEproof}
$p_m(\tilde s)$ is given by
\begin{equation}
  p_m(\tilde s)=\int_0^1 (1-F(\tilde s x))\d x=\int_0^1 \frac{\Gamma(m,m\tilde sx)}{\Gamma(m)}\d x\,.
\end{equation}
For $m\in\mathbb{N}$, this is
\begin{equation}
  p_m(\tilde s)=\sum_{k=0}^{m-1} \int_0^1 \exp(-m\tilde s x)\frac{(m\tilde s x)^k}{k!}\d x\,,
\end{equation}
which, after some manipulations, yields
\begin{align}
 p_m(\tilde s)&=\frac 1{\tilde s}\left(1-\frac 1 m \exp(-m\tilde s)\sum_{k=0}^{m-1}\sum_{j=0}^k 
    \frac{(m\tilde s)^j}{j!}\right)\,\\
    &=\frac 1{\tilde s}\left(1-\exp(-m\tilde s)\underbrace{\sum_{k=0}^{m-1} \frac{m^k(1-k/m)}{k!} \tilde{s}^k}_{P_{m-1}(\tilde s)}\right)\,.
\end{align}
The polynomial $P_{m-1}$ is the Taylor expansion
of order $m$ of $(1-\tilde s)\exp(m\tilde s)$ at $\tilde s=0$ (the coefficient for $\tilde{s}^m$ is zero).
So $\exp(-m\tilde s) P_{m-1}(\tilde s)=1-s+O(s^{m+1})$ from which the limit $1$ for $\tilde s<1$ follows.
For $\tilde s>1$, the exponential dominates the polynomial so that their product tends
to zero and $1/\tilde s$ remains as the limit.
\end{IEEEproof}

The convergence to $\min\{1,\tilde{s}^{-1}\}$ is  the expected behavior,
since without fading a node is connected if it is positioned within $[0,1/s]$ ($\tilde s<1$)
and for a randomly
chosen node in $[0,a]$ for $a>1/s$ or $\tilde s>1$, this has probability $1/as$. So with increasing $m$,
derivatives of higher and higher order become 0 at $\tilde s=0$. From the previous discussion
we know that $p_m(\tilde s)=1+O(\tilde{s}^m)$. Calculating the coefficient for $\tilde{s}^m$
yields
\begin{equation}
  p_m(\tilde s)=1-\frac{m^m}{\Gamma(m+2)} \tilde{s}^m + O(\tilde{s}^{m+1})\,.
  \label{taylor}
\end{equation}
The $m$-th order Taylor expansion at $\tilde s=0$ is a lower bound.
Upper bounds are obtained by truncating the
polynomial; a natural choice is the first-order version $1+(m-1)\tilde s$
to obtain
\begin{equation}
  \left(1-\frac{m^m}{\Gamma(m+2)} \tilde{s}^m\right)^+ < p_m(\tilde s) \leqslant
    \min\left\{1, \frac 1{\tilde s}\left(1-\exp(-m\tilde s)(1+(m-1)\tilde s)\right)\right\}\,.
    \label{low_bound}
\end{equation}

\noindent Using the lower bound, we can establish the following Corollary.

\begin{corollary}[$\epsilon$-reachability.]
If
\begin{equation}
 as < \frac{\left(\Gamma(m+2)\cdot\epsilon\right)^{1/m}}{m}\,.
 \label{broad_as_gen}
\end{equation}
 at least a fraction $1-\epsilon$ of the nodes $\x_i\in [0,a)$
 are connected.
  In the standard network (specializing to $m=1$), the sufficient
  condition is 
  \begin{equation}
  as<2\epsilon\,,
  \label{broad_as}
\end{equation}
\label{cor:broad}  
 \end{corollary}
 This follows directly from the lower bound in \eqref{low_bound}.
 
\noindent{\em Remarks:}
\begin{itemize}
\item[-] For $m\rightarrow\infty$, the bound \eqref{broad_as_gen} is not tight since the RHS converges
to $1/e$ for all positive $\epsilon$ (by Stirling's approximation), while the
exact condition is $as<1/(1-\epsilon)$.
\item[-] The sufficient condition \eqref{broad_as} is tight (within 7\%) for $\epsilon<0.1$.
 With $p_1(as)=(1-e^{-as})/as$, the condition $p_1(as)>1-\epsilon$
can be solved exactly using the Lambert W function: 
\begin{equation}
 as < \W(-q e^{-q})+q\,,\qquad \text{where }q\triangleq \frac{1}{1-\epsilon}\,.
\end{equation}
A linear approximation yields the same bound
as before, while a quadratic expansion yields the sufficient condition
 $as<2\epsilon+4/3\epsilon^2$ which is
within $3.9\%$ for $\epsilon< 0.25$.
\end{itemize}

\subsection{Broadcast transport sum-distance and capacity}
Assuming the origin $o$ transmits, the set of nodes
that receive the message is $\{\hat\x_i\}$. We shall
determine the {\em broadcast transport sum-distance} $D$, \ie,
the expected sum over the all the distances $\hat\x_i^{1/\alpha}$
from the origin:
\begin{equation}
D\triangleq\E\left(\sum_{\x\in\hat\Phi} \x^{1/\alpha}\right)
\end{equation}

\begin{proposition}
\label{prop:btd}
The broadcast transport sum-distance for Nakagami-$m$ fading is
\begin{equation}
  D_m=c_d\frac\delta\Delta \frac1{(ms)^\Delta}\frac{\Gamma(m+\Delta)}{\Gamma(m)}\,,
  \label{ct}
\end{equation}
and the (broadcast) {\em fading gain} $D_m/D_{\infty}$ is
\begin{equation}
\frac{D_m}{D_\infty}=
  \frac 1{m^\Delta}\frac{\Gamma(m+\Delta)}{\Gamma(m)}=\E(\f^\Delta)\,.
  \label{fading-gain}
\end{equation}
\end{proposition}
\begin{IEEEproof}
From Campbell's theorem 
\begin{align*}
\E\left(\sum_{\x\in\hat\Phi} \x^{1/\alpha}\right)&=\int_0^\infty x^{1/\alpha} \hat\lambda(x)\d x \\
  &=c_d\delta \int_0^\infty x^{1/\alpha+\delta-1}(1-F(sx))\d x\,,
\end{align*}
which equals \eqref{ct} for Nakagami-$m$ fading.

Without fading, a node $\x_i$ is connected if $\x_i<1/s$,
therefore
\begin{align}
  D_{\infty} &=\int_0^{1/s} x^{1/\alpha}\lambda(x)\d x\\
     &= c_d \frac\delta\Delta s^{-\Delta}=c_d \frac{d}{d+1} s^{-\Delta}\,.
\end{align}
So the fading gain $D_m/D_\infty$ is the $\Delta$-th moment
of $\f$ as given in \eqref{fading-gain}.
\end{IEEEproof}

\noindent{\em Remarks:}
\begin{enumerate}
\item The fading gain is independent of the threshold $s$.
  $D_m\propto s^{-\Delta}$ for all $m$. It strongly resembles
  the connectivity gain (Prop.~\ref{prop:con}), the only
  difference being the parameter $\Delta$ instead of $\delta$.
  In particular, $D_m$ is independent of $m$ if $\Delta=1$.
  See Remark 3 to Prop.~\ref{prop:con} and \figref{fig:gains} for a
  discussion and visualization of the
  behavior of the gain as a function of $m$ and $\Delta$.
\item For Rayleigh fading ($m=1$), $D_1=c_d\delta s^{-\Delta}$, and
the fading gain is $\Gamma(1+\Delta)$.
For $d=\alpha=2$, $D_{\infty}=\frac{2\pi}{3s^{3/2}}$.
\item
The formula for the broadcast transport sum-distance reminds of an
interference expression. Indeed, by simply replacing $\x^{1/\alpha}$
by $\x^{-1}$, a well-known result on the mean interference is reproduced:
Assuming each node transmits at unit power,
the total interference at the origin is
 \[  \E\left(\sum_{\x\in\Phi} \x^{-1}\right)=\int_0^\infty x^{-1} \lambda(x)\d x=
  c_d\frac\delta{\delta-1}x^{\delta-1}\Big|_0^\infty \]
which for $\delta<1$ diverges due to the lower bound integration
bound (\ie, the one or two closest nodes) and
for $\delta\geqslant 1$ diverges due to the upper bound (\ie, the
large number of nodes that are far away).
\end{enumerate}
 
So far, we have ignored the actual rate of transmission $R$ and just
used the threshold $s$ for the sum-distance. To get to the single-hop
broadcast transport capacity $C$ (in bit-meters/s/Hz), we relate the
(bandwidth-normalized) rate of transmission $R$ and the threshold $s$
by $R=\log_2(1+s)$ and define
\begin{equation}
  C\triangleq \max_{R>0} \{R\cdot D(2^R-1)\}=\max_{s>0} \{\log_2(1+s) D(s)\}.
\end{equation}
Let $D_m^1$ be the broadcast transport sum-distance for $s=1$
(see Prop.~\ref{prop:btd}) such that $D_m=D_m^1 s^{-\Delta}$.

\begin{proposition}
For Nakagami-$m$ fading:
\begin{enumerate}[(a)]
\item For $\Delta\in (0,1)$, the broadcast transport capacity is
achieved for 
\begin{equation}
  \Ropt=\frac{\W\left(-\frac{e^{-1/\Delta}}\Delta\right)+\Delta^{-1}}{\log 2}\,,\quad
  \Delta\in (0,1)\,.
\label{ropt}
\end{equation}
The resulting broadcast transport capacity is tightly (within at most
0.13\%) lower bounded by
\begin{equation}
 C_m \geqslant \frac{D_m^1(\Delta)}{\log 2}(\Delta^{-1}-\Delta)\left(e^{\Delta^{-1}-\Delta}
-1\right)^{-\Delta}\,.
\label{cmbound}
\end{equation}
\item For $\Delta=1$,
\begin{equation}
   C_m=\frac{c_d\delta}{\log 2}\,
\end{equation}
 independent of $m$, and $R_{\text{opt}}=0$.
\item For $\Delta>1$, the broadcast transport
capacity increases without bounds as $R\rightarrow 0$, independent of
the transmit power.
\end{enumerate}
\end{proposition}
\begin{IEEEproof}
\begin{enumerate}[(a)]
\item $D_m\propto s^{-\Delta}$, so $C_m\propto R(2^R-1)^{-\Delta}$ which, for
$\Delta\leqslant 1$, has a maximum at $R_{\mathrm{opt}}$ given in \eqref{ropt}.
The lower bound stems from an approximation of $R_{\mathrm{opt}}$ using
$\W(-\exp(-1/\Delta)/\Delta)\gtrapprox -\Delta$ which holds since for $\Delta=1$, the two expressions
are identical, and the derivative of the Lambert W expression is smaller than -1
for $\Delta<1$. 
\item For $\Delta=1$, $C_m$ increases as the rate is lowered but remains bounded
 as $R\rightarrow 0$. The limit is $c_d\delta/\log 2$.
\item For $\Delta>1$, $R(2^R-1)^{-\Delta}$ is decreasing with $R$, and
$\lim_{R\rightarrow 0} R(2^R\!-\!1)^{-\Delta}=\lim_{R\rightarrow 0}
 (\log 2)^{-\Delta} R^{1-\Delta}=\infty$. 
\end{enumerate}
\end{IEEEproof}
{\em Remarks:}
\begin{itemize}
\item[-] The optima for $R$, $s$ are independent of the type of fading (parameter $m$).
\item[-] For $\Delta<1$, the optimum $s$ is tightly lower bounded by
\begin{equation}
  s_{\text{opt}} \geqslant \exp(\Delta^{-1}-\Delta)-1\,.
\end{equation}
 This is the expression appearing in the bound \eqref{cmbound}.
\item[-] (c) is also apparent from the
 expression $D(s)\log_2(1+s)$, which, for $s\rightarrow 0$, is
 approximately $D_m^1 s^{1-\Delta}/\log 2$. So, the intuition is that 
 in this regime, the
 gain from reaching additional nodes more than offsets the loss in rate.
 \item[-]  For $\Delta=1/(2\log 2)$, $s_{\text{opt}}=R_{\text{opt}}=1$ and $C_m=D_m^1$. This is, however, not the minimum. The capacity is minimum around $\Delta\approx 0.85$,
   depending slightly on $m$.
 \end{itemize}

\begin{figure}
\centerline{\epsfig{file=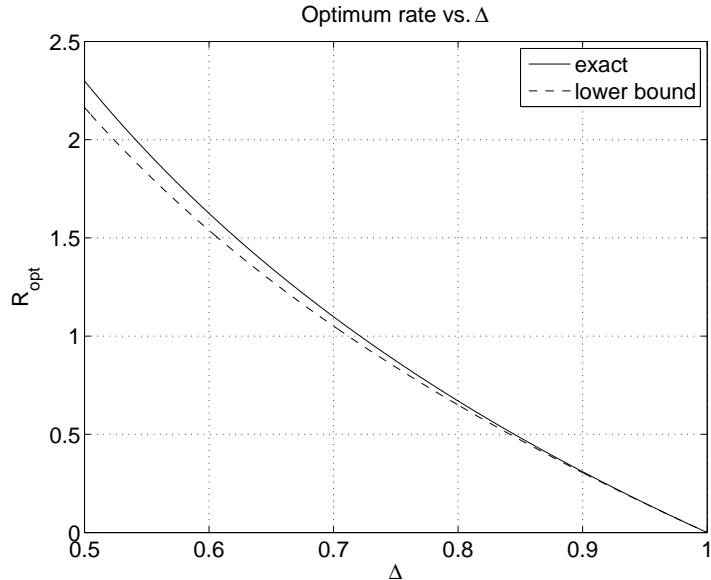,width=\figwidth}}
\caption{Optimum transmission rates for $\Delta\in [0.5,1.0]$ The optimum rate is $1$ for $\Delta=1/(2\log 2)\approx 0.72$.}
\label{fig:opt_rates}
\end{figure}

\begin{figure}
\centerline{\epsfig{file=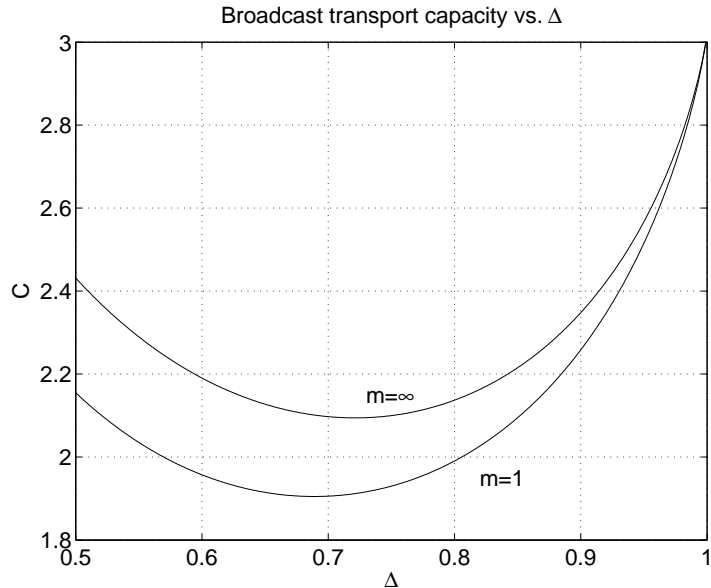,width=\figwidth}}
\caption{Broadcast transport capacity for $d=2$, $\Delta\in [0.5,1.0]$ and $m=1$ and $m=\infty$. For
$\Delta=1$, the capacity is $2\pi /(3\log 2)\approx 3.02$ irrespective of $m$. For the no
fading case, the minimum occurs at $\Delta=1/(2\log 2)$, where $C=2\pi/3$.}
\label{fig:trsp_capacity}
\end{figure}

\figref{fig:opt_rates} depicts the optimum rate as a function of $\Delta$, together
with the lower bound $(\Delta^{-1}-\Delta)/\log 2$, and \figref{fig:trsp_capacity}
plots the broadcast transport capacity for Rayleigh fading and no fading
for a two-dimensional network. The range $\Delta\in[0.5,1.0]$ corresponds
to a path loss exponent range $\alpha\in[3,6]$. It can be seen that Nakagami fading
is harmful. For small values of $\Delta$, the capacity for Rayleigh fading is
about 10\% smaller.

\subsection{Optimum broadcasting (superposition coding)}
Assuming that nodes can decode at a rate corresponding to their
SNR, the broadcast transport capacity (without fading) is
\begin{equation}
  \tilde C=\E\left[\sum_{\x\in\Phi} \x^{1/\alpha} \log_2(1+\x^{-1})\right] 
\end{equation}
To avoid problems with the singularity of the path loss law at the origin, we
replace the $\log$ by 1 for $x<1$. For $x>1$, we use the lower bound $\log_2(1+x^{-1})>1/x$.
Proceeding as in the proof of Prop.~\ref{prop:btd}, we obtain
\begin{equation}
  \tilde C>c_d\delta \left(\frac 1\Delta +\int_1^\infty x^{\Delta-2}\d x\right)\,,
\end{equation}
which is significantly larger than in the case with single-rate decoding.
For $\Delta<1$, 
\begin{equation}
\tilde C>\frac{c_d\delta}{\Delta(1-\Delta)}\,.
\end{equation}
For $\Delta\geqslant 1$, this lower bound and thus $\tilde C$ is unbounded, in
agreement with the previous result. The only difference is that for
$\Delta=1$, $\tilde C$ diverges whereas $C$ is finite.
Note that since $\log_2(1+x^{-1})<1/(x\log 2)$ for $x>1$, the lower bound is within a factor
$\log 2$ of the correct value.

If the actual Shannon capacity were considered for nodes that are
very close, $\tilde C$ would diverge more quickly as $\Delta\rightarrow 0$ 
($\alpha\rightarrow\infty$) since the contribution from the nodes within
distance one would be:
\begin{equation}
 \tilde C_{[0,1]}> c_d\delta \int_0^1 -x^{\Delta-1}\log_2 x\, \d x=\frac 1{\log(2)\Delta^2}\,.
\end{equation}

\section{Other Applications}

\subsection{Maximum transmission distance}
How far can we expect to transmit, \ie, what is the
(average) {\em maximum transmission distance}
$M\triangleq\E\left(\max_{\x\in\hat\Phi}\{\x^{1/\alpha}\}\right)$?

\begin{figure}
\centerline{\epsfig{file=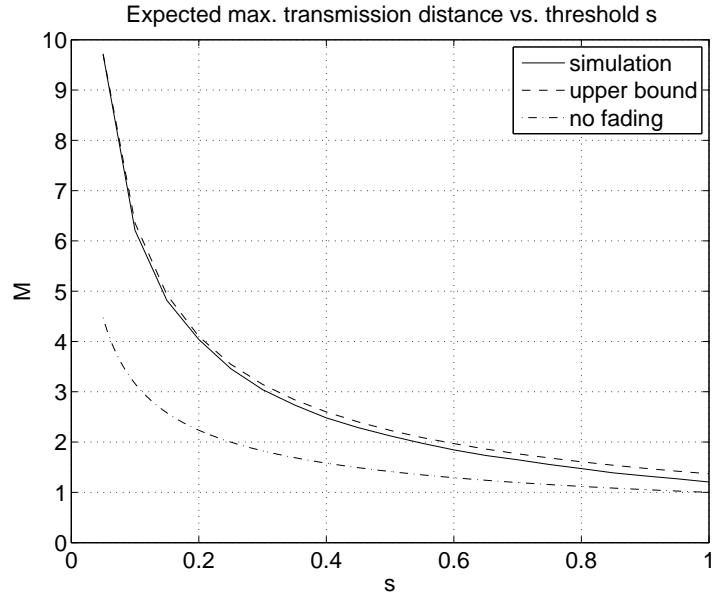,width=\figwidth}}
\caption{Expected maximum transmission distances for the standard two-dimensional network
and $s\in [0.05,1.00]$. For comparison, the curve $s^{-1/2}$ for the non-fading case is also
displayed.}
\label{fig:gain}
\end{figure}

Let $\hat\x$ be a uniformly randomly chosen connected node.
The pdf $f_{\hat\x}$ is given by \eqref{randomconnected_pdf}.
The distribution of the maximum $\x_M$ of a Poisson number of RVs is
given by the Gumbel distribution\footnote{Note that the Gumbel cdf is
not zero at $0^+$. This reflects the fact that the number of connected nodes
may be zero, in which case the maximum transmission distance would
be zero. Accordingly the pdf includes a pulse at $0$, the term $\exp(-\E\hat N)\delta(x)$.}
\begin{equation}
  F_{\hat\x_M}(x)=\exp\left(-\E\hat N (1-F_{\hat\x}(x)\right)\,.
\end{equation}
So, in principle, $M=\E(\hat\x_M^{1/\alpha})$ can be calculated.
However,
even for the standard network, where
$F_{\hat x_M}(x)=\exp(-\frac{c_d}s \exp(-sx))$, there does not seem to exist
a closed-form expression.
If the number of connected nodes was fixed to $c_d/s$ (instead of being Poisson
distributed with this mean), we would have $F_{\hat x_M}(x)=(1-e^{-xs})^{c_d/s}$
with mean
\begin{equation}
  \E\hat \x_M=\frac 1 s \left(\Psi\left(\frac{c_d}{s}+1\right)+\gamma\right)\,.
\end{equation}
Since $\Psi$ is concave, this upperbounds the true mean by Jensen's inequality.
Finally, we invoke Jensen again by replacing $\E(\hat\x_M^{1/\alpha})$ by
$\E(\hat\x_M)^{1/\alpha}$ to obtain 
\begin{equation}
  M < \left(\frac 1 s \left(\Psi\left(\frac{c_d}{s}+1\right)+\gamma\right)\right)^{1/\alpha}\,.
\end{equation}
Without much harm, $\Psi(x)$ could be replaced by (the slightly larger) $\log(x)$. Even
replacing $\Psi(x+1)$ by $\log(x)$ still appears to be an upper bound.
The bound is quite tight,
see \figref{fig:gain}. Also compare with \figref{fig:disks2}, where the
most distant node is quite exactly 6 units away ($s=0.1$).
The factor $s^{-1/\alpha}$ is the bound in the non-fading case, so the
Rayleigh fading (diversity) gain for the maximum transmission distance is roughly
$\log(1/s)^{1/\alpha}$ which grows without bounds as $s\rightarrow 0$.

\subsection{Probabilistic progress}
In addition to the maximum transmission distance or 
the distance-rate product, the product distances times probability
of success may be of interest. 
Without considering the actual node positions, one may want to maximize
the {\em continuous probabilistic progress}
$G(x)\triangleq\max\{x^{1/\alpha}\P[\f>sx]\}$.
For the standard network with $\alpha=2$, this is maximized at
$x=1/2s$. If there was no fading, the optimum would be $x=\sqrt{1/s}$.
Of course there is no guarantee that there is a node very close to this
optimum location.

Alternatively, define the {\em (discrete) probabilistic progress} when transmitting to node $i$ by
\begin{equation}
 G_i\triangleq\E\left(\x_i^{1/\alpha} \cdot \P[\f> s\x_i\,|\,\x_i]\right) 
\end{equation}
We would like to find $i_\opt=\argmax_i G_i$.
For the standard network,
\begin{equation}
 G_i=\E\left(\x_i^{1/\alpha} \exp(-s\x_i)\right)=\frac{c_d^i}{(s+c_d)^{i+1/\alpha}}\frac{\Gamma(i+1/\alpha)}
 {\Gamma(i)}\,.
 \label{gi}
\end{equation}
The maximum of $G_i$ cannot be found directly, but since
$\Gamma(i+1/\alpha)/\Gamma(i)$ is very tightly lower bounded by $i^{1/\alpha}$ we have
\begin{equation}
 G_i\lessapprox \frac{c_d^i\, i^{1/\alpha}}{(s+c_d)^{i+1/\alpha}}
\end{equation}
which, assuming a continuous parameter $\tilde i$, is maximized at
\begin{equation}
  \tilde i_{\opt}=\frac 1{\alpha\log(1+\frac{s}{c_d})}\,.
\end{equation}

Note that the same expression for $i_\opt$ would be obtained if $G_i$ was approximated
by the factorization $G_i'=\E(\x_i^{1/\alpha})\P[\xi_i<1/s]$. For the standard network,
$\E(\x_i^{1/\alpha})=\frac{\Gamma(i+1/\alpha)}{\Gamma(i)\,c_d^{1/\alpha}}$, and
$\P[\xi_i<1/s]=(\pi/(\pi+s))^i$. So $G_i'$ differs from $G_i$ only by the
factor $(1+s/c_d)^{1/\alpha}$ which is independent of $i$ and quite small
for typical $s$.

Now, the question is how to round $\tilde i_\opt$ to $i_\opt$. For large $s$,
$i_\opt=1$. For small $s$, $\tilde i_\opt\approx c_d/(\alpha s)$ so
\begin{equation}
  i_\opt=\lceil \frac{c_d}{\alpha s}\rceil
\end{equation}
is a good choice. It can be verified that this is indeed the optimum.
The expected distance to this $i_\opt$-th node is quite exactly
$1/(\alpha s)^{1/\alpha}$. So in this non-opportunistic setting when
reliability matters, Rayleigh fading is harmful; it reduces the range of transmissions
by a factor $\alpha^{-1/\alpha}$.

\subsection{Retransmissions and localization}
\begin{proposition}[Retransmissions]
Consider a network with block Rayleigh fading. The expected number of
nodes that receive $k$ out of $n$ transmitted packets $\E N_k^n$ is
\begin{equation}
\E N_k^n=\frac{c_d\Gamma(1+\delta)}{(ks)^\delta}\,,\quad k\in\{0,1,\ldots,n\}\,.
\label{enkn}
\end{equation}
\label{prop:retrans}
\end{proposition}
\begin{IEEEproof}
Let $p(x)\triangleq 1-F(sx)$.
The density of nodes that receive
$k$ packets out of $n$ transmissions is given by
\begin{equation}
  \lambda_k^n(x)=\lambda(x) \binom n k p(x)^k (1-p(x))^{n-k}\,.
\label{retrans-density}
\end{equation}
Plugging in $p(x)=\exp(-sx)$ for Rayleigh fading and 
integrating \eqref{retrans-density} yields $\E N_k^n=\Lambda_k^n(\mathbb{R}^+)$.
\end{IEEEproof}
\noindent{\em Remarks:}
\begin{enumerate}[-]
\item Interestingly, \eqref{enkn} is independent of $n$. So, the mean number of nodes
  that receive $k$ packets does not depend on how often the packet was transmitted.
\item Summing $\lambda_k^n$ over $k\in[n]$ reproduces Cor.~\ref{cor:retrans}.
\item \eqref{enkn} is valid even for $k=0$ since $\E N_0^n=\infty$.
\item For the standard networks, the expression simplifies to
$\E N_k^n=\frac{c_d}{ks}$,
which, when summed over $k\in[n]$, yields \eqref{conn_n}. 
\end{enumerate}

Let $\x_k^n$ be the position of a randomly chosen node from the nodes that
received $k$ out of $n$ packets. From Prop.~\ref{prop:retrans}, the pdf (normalized
density) is
\begin{equation}
    f_{\x_k^n}(x)=\lambda_k^n(x)\frac{(ks)^\delta}{c_d\Gamma(1+\delta)}\,,\quad k\in [n]\,.
    \label{kn_density}
\end{equation}

For the standard network, we have $\E\x_n^n=(ns)^{-1}$, $\V\x_n^n=(ns)^{-2}$, and $\E\x_1^n=\frac1s(\Psi(n+1)+\gamma)$,
which is again related to \eqref{conn_n} (division by the constant density $c_d$).

\begin{figure}
\centerline{\epsfig{file=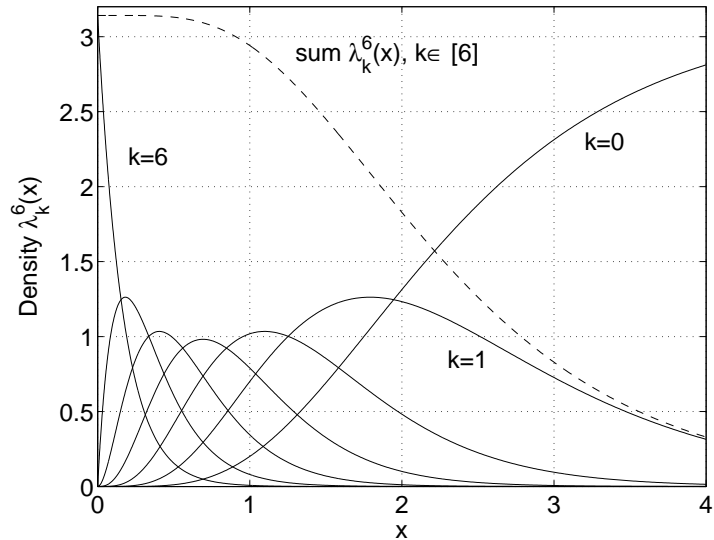,width=\figwidth}}
\caption{Densities $\lambda_k^6(x)$ for the standard network with $\alpha=2$ ($c_d=\pi$)
and $s=1$. The maximum of the density for $k=n=6$ is $\lambda_6^6(0)=\pi$. The dashed curve is
  the density of the nodes that receive at least 1 packet. Normalized by $\E N_k^6$ these densities
  are the pdfs of $\x_k^6$.}
  \label{fig:densities}
\end{figure}

The densities of the nodes receiving exactly $k$ of
$6$ messages is plotted in
\figref{fig:densities} for the standard network with $\alpha=2$.

This expression permits the evaluation of the contribution that each
additional transmission makes to the broadcast transport sum-distance and
capacity.

These results can also be applied in localization. If a node receives
$k$ out of $n$ transmissions, $\E \x_k^n$ is an obvious estimate for its position,
and $\V\x_k^n$ for the uncertainty.
Alternatively, if the path loss $x$ can be measured, then the corresponding node
index $\hat i(x)$ can be determined by the ML estimate
\begin{equation}
  \hat{i}(x)=\argmax_i f_{\xi_i}(x)\,,
\end{equation}
with the pdf $f_{\xi_i}$ given in Cor.~\ref{cor:basic}. For the standard networks, for example,
the ML decision is $\hat{i}(x)=\lceil c_d/x \rceil$ since
\begin{equation}
 \hat{i}(x)=i\quad\Longleftrightarrow \quad\frac{c_d}{i} \leqslant x < \frac{c_d}{i-1}\,.
\label{ml_bounds}
\end{equation}
This is of course related to the fact $\E\x_i=i/c_d$.

\section{Concluding Remarks}
We have offered a geometric interpretation of fading in wireless networks
which is based on a point process model that incorporates both geometry and fading.
The framework enables analytical investigations of the properties of wireless networks and the
impact of fading, leading to closed-form results that are obtained in a rather convenient manner.

For Nakagami-$m$ fading, it turns out that the {\em connectivity
fading gain} is the $\delta$-th moment of the fading distribution,
while the fading gain in the {\em broadcast transport sum-distance} is
its $\Delta$-th moment.
A path loss exponent larger than the
number of dimensions $d$ ($d+1$ for broadcasting) leads to
a negative impact of fading.
Interestingly, the {\em broadcast transport capacity} turns out to be
unbounded if $\Delta>1$, \ie, if the path loss exponent is smaller
than $d+1$. While this result may be of interest for the design of
efficient broadcasting protocols, it also raises doubts on the
validity of transport capacity as a performance metric.

Generally, it can be observed that the parameters $\delta$ and/or $\Delta$
appear ubiquitously in the expressions. So the network behavior critically
depends on the ratio of the number of dimensions to the path loss exponent.

Other applications considered include the maximum transmission distance,
probabilistic progress, and the effect of retransmissions. We are convinced
that there are many more that will benefit from the theoretical foundations
laid in this paper.

\section*{Acknowledgments}
The support of NSF (Grants CNS 04-47869, DMS 505624) and the DARPA IT-MANET
program (Grant W911NF-07-1-0028) is gratefully acknowledged.

\bibliographystyle{IEEEtr} 
\bibliography{header,comm,net}

\end{document}